\documentclass[psfig]{aipproc}

\layoutstyle{8x11double}

\usepackage{psfig}

\SetInternalRegister\hbadness{8000} 

\begin{document}

\title 
      [Neutron Star Binaries as Central Engines of GRBs]
      {Neutron Star Binaries as Central Engines of GRBs}


\author{S. Rosswog}{
  address={Dep. Phys. and Astronomy, University of Leicester, 
 LE1 7RH Leicester, UK},
  email={sro@star.le.ac.uk},
}

\copyrightyear  {2001}

\begin{abstract}
We describe the results high resolution, hydrodynamic calculations of 
neutron star mergers. The model makes use of a new, nuclear equation of state,
accounts for multi-flavour neutrino emission and solves the equations of 
hydrodynamics using the smoothed particle hydrodynamics method with more than
$10^6$ particles. The merger leaves behind a strongly differentially rotating
central object of $\sim 2.5$ M$_{\odot}$ together with a distribution of
hot debris material. For the most realistic case of initial neutron star spins,
no sign of a collapse to a black hole can be seen. We argue that the 
differential rotation stabilizes the central object for $\sim 10^2$ s 
and leads to superstrong magnetic fields.
We find the neutrino emission from the hot debris around the freshly-formed,
supermassive neutron star to be substantially lower than predicted previously.
Therefore the annihilation of neutrino anti-neutrino pairs will have 
difficulties to power very energetic bursts ($\gg 10^{49}$ erg).

\end{abstract}

\date{\today}

\maketitle

\section{Introduction}

There is growing observational evidence that the subclass of long 
Gamma Ray Bursts (GRBs) is related to star forming regions 
(e.g. \cite{bloom02}). 
While this connection is immediately evident for 
the short-lived progenitors of collapsars, the question of whether neutron 
star binaries merge close to star forming regions or not is not a settled one.
Bloom et al. (2002), for example, argue against compact object mergers as 
central engines of (at least the subclass of the long) GRBs.
Belczynski et al. (2001), however, claim to have identified new formation
channels that lead to classes of very tight, short-lived neutron star 
binaries. They find typical inspiral times of the order $10^6$ years and 
therefore neutron star binaries would merge very close to their birth places, 
even if equipped with high systemic velocities due to kicks in asymmetric 
supernova explosions.\\
Recent afterglow observations of long bursts suggest that they are beamed
and require $E_\gamma \sim 5 \cdot 10^{50}$ erg \cite{frail01}. The energy 
requirements for the subclass of short bursts are so far essentially 
unconstrained.
Due to their gravitational binding energy of several times $10^{53}$ erg, 
neutron star binaries certainly do possess the energy reservoirs necessary
to power a (long) burst, however, how to transform
the available energy into gamma rays is still far from being clear. Among 
the suggested mechanisms are magnetic energy extraction processes  (e.g.
\cite{duncan92,narayan92,usov94}) and the annihilation 
of neutrino-antineutrino pairs emitted from the hot neutron star debris 
during the coalescence \cite{eichler89}.

\section{High Resolution Simulations of the Merger Event} 
To study the possible role of neutron star coalescences for either long or 
short bursts we have performed detailed high-resolution simulations of the 
merger event \cite{rosswog01a}. To solve the equations of fluid dynamics we 
apply the smoothed 
particle hydrodynamics method (SPH) together with a largely improved
artificial viscosity tensor \cite{rosswog00}. We treat the self-gravity 
of the neutron star fluid in a 
Newtonian way, but we add the forces emerging from the emission of 
gravitational waves to drive the system towards coalescence. The microphysical
properties of the hot and dense neutron star matter are described using an 
equation of state (EOS) that is based on the tables of Shen et al. 
\cite{shen98a,shen98b}. We have extended the EOS to the low density regime via a gas 
consisting of 
neutrons, protons, alpha particles, electron-positron pairs and photons. 
This new EOS covers the whole relevant parameter space in density, temperature
and electron fraction. To allow for cooling and compositional changes, we 
have implemented a detailed neutrino treatment that accounts for three 
neutrino flavours ($\nu_e$,$\bar{\nu}_e$, and $\nu_x$, which is collectively 
used for the four
heavy-lepton neutrinos) and takes the relevant emission processes (lepton 
capture on nucleons, electron-positron pair annihilation and plasmon decay) 
into account. The opacities are calculated from the absorption processes 
of the 
electron-type neutrinos on nucleons and scattering off nucleons and nuclei,
the latter process becoming the dominant opacity source even for moderate 
mass fractions of heavy nuclei. The calculations are performed efficiently
on shared-memory parallel computers with more than $10^{6}$ SPH-particles. 
For a more detailed description of the input physics and the computational 
methods we refer to Rosswog et al. \cite{rosswog01a,rosswog01b}.

\section{Results}
\begin{figure}
\psfig{file=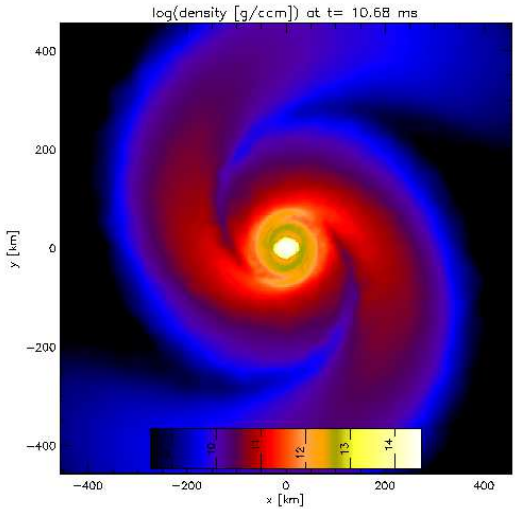,width=7cm,angle=0}
\label{nice}
\caption{Coalescence of a corotating binary system of 1.4 M$_{\odot}$ per 
star. Colour-coded is the matter density in the orbital plane. More than 
$10^6$ SPH-particles were used for this calculation.}
\end{figure}
In this latest set of calculations we explore systems with two initial spin 
configurations (for details see \cite{rosswog01a}): tidally locked, corotating
 systems for the ease of constructing initial equilibrium
models and systems without initial neutron star spins. The latter ones are
the most relevant spin configurations since the intrinsic neutron 
star viscosity
is too low to lead to a tidal locking during the short phase where the 
binary components undergo tidal interaction \cite{bildsten92,kochanek92}.
We find central objects of 2.3 to 2.6 M$_{\odot}$ which are strongly 
differentially rotating. For the most realistic, irrotational case the 
maximum density does not even reach the initial density of a single, cold
non-rotating neutron star.
Apart from the thermal pressure and the differential rotation there may be 
further effects that, at least temporarily, stabilize the merger remnant 
against collapse to a black hole: the presence of non-leptonic, negative 
charges together with trapped neutrinos \cite{prakash95}, for example, can 
substantially increase the maximum possible mass. One also expects magnetic 
seed fields to be amplified in the differentially rotating remnant to enormous
field strengths ($\sim 10^{17}$ G) \cite{duncan92}. Fields of this strength
can substantially modify the structure of the central object and provide 
additional support against collapse \cite{cardall01}.
Due to our ignorance of the high-density equation of state it cannot be 
ruled out that the end product of the coalescence is a stable supermassive 
neutron star of $\sim$ 2.8 M$_{\odot}$. It seems, however, more likely
that the central object is only temporarily stabilized and once the stabilizing
effects weaken (e.g. neutrinos have diffused out after $\sim 10$ s, 
magnetic braking has damped out differential rotation) collapse to a 
black hole will set in. The time scale until collapse is difficult to 
determine, since it depends sensitively on poorly known physics and on the 
specific system parameters. We expect the most important effect to come from
the differential rotation \cite{ostriker68,baumgarte00} and therefore the
collapse time scale to be set by the time it takes the remnant to reach 
uniform rotation. Assuming the dominant 
effect to come from magnetic dipole radiation (the viscous time scale is estimated to be $\sim 10^9$ s \cite{shapiro00}), this time is given by
\begin{eqnarray}
\tau_c&\sim&\frac{18 c^3 M}{5 B^2 R^4 \omega^2} \nonumber\\ 
&\sim&10^2 s \left(\frac{M}{2.5 M_\odot} \right) \left( \frac{10^{16} G}{B} 
\right)^2
\left( \frac{15 km}{R} \right)^4 \left( \frac{3000 s^{-1}}{\omega} \right)^2,
\nonumber
\end{eqnarray}
where $M, B$ and $R$ are mass, magnetic field and radius of the central object.
Such an object, an at least temporarily stabilized, 
supermassive neutron star with enormous magnetic field strength, is at the 
heart of many suggested GRB models (e.g. \cite{duncan92,usov94,kluzniak98}).
Kluzniak and Ruderman, for example, estimate that the magnetic field becomes 
buoyant at $\sim 10^{17}$ G, floats up and breaks through the surface of the
remnant as a sub-burst. This process, winding up the field to buoyancy, 
subsequent floating up and sub-burst, would continue until the energy stored 
in differential 
rotation is used up (or collapse sets in), $\sim$ 100 s. Therefore even long 
bursts, with substructures on millisecond time scale set by the motion of 
the magnetized fluid, could result from a neutron star merger. Once the 
differential rotation has
been damped out, the system could still continue as a gamma ray burster using
the energy stored in rigid rotation \cite{usov92} in case it remains 
stable, or, in case of collapse, one would be left with the 'classic'
GRB-engine, a black hole plus a debris torus.\\
\begin{figure}
\psfig{file=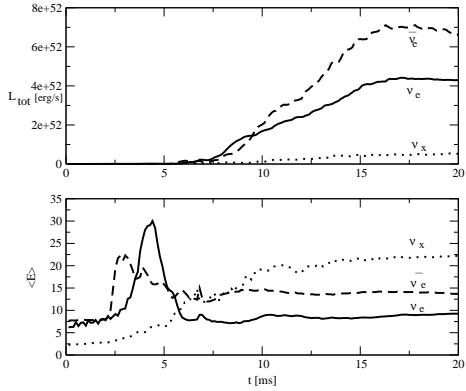,width=7.5cm,angle=-90}
\label{nu_lum}
\caption{Total neutrino luminosities and mean energies (in MeV) 
for the $\nu_e$, 
 $\bar{\nu}_e$ and $\nu_x$ emitted in the coalescence of a corotating neutron
 star binary system of twice 1.4 M$_{\odot}$. The abscissa gives the time in 
ms.}
\end{figure}
The annihilation of neutrino anti-neutrino pairs into electron-positron pairs
above the poles of the merger remnant has been suggested \cite{eichler89}
as a process to produce a fireball from thermal energy stored in the disk.
We find that the main neutrino-emitting region is the inner, shock- and 
shear-heated region of the torus around the central object.
The prevailing densities lie between $10^{10}$ und 
$10^{12}$ gcm$^{-3}$ and temperatures are $\sim $ 2-3 MeV.
In our simulations we find typical neutrino
luminosities of $\sim 10^{53}$ ergs with mean energies of $\sim 10, \sim 15$ 
and $\sim 20$ MeV (see Fig. \ref{nu_lum}) for an initially corotating system, 
and slightly higher values for irrotational systems \cite{rosswog01b}. The dominant emission process is the capture of 
positrons onto neutrons which occurs in the hot, neutron-rich inner regions 
of the debris torus.
The found luminosities are lower than those found previously by 
Ruffert and Janka \cite{ruffert01} by roughly a factor of 3 to 4.
The annihilation efficiency is currently being investigated in detail, but the 
preliminary results indicate that it will be difficult to power Gamma Ray 
Bursts largely in excess of $\sim 10^{49}$ ergs.

\section{Summary}

We have performed high resolution calculations of neutron star coalescences 
with more than $10^6$ SPH-particles, using a new, nuclear equation of state
and accounting for the effects of cooling and compositional changes from 
neutrino reactions by a detailed multi-flavour neutrino treatment.
For the most realistic case with negligible neutron star spins, we do not see
any sign of a collapse to a black hole and therefore argue that the outcome
of the coalescence is a (at least temporarily) stabilized, supermassive, hot, 
differentially rotating object with huge magnetic fields. Such an object could 
produce a GRB in various ways, e.g. via a relativistic electron-positron
wind (Usov 1994) or via the so-called DROCO-mechanism (differentially rotating
compact object) suggested by Kluzniak and Ruderman (1998)
where the field inside the differentially rotating central object becomes 
locally wound up until it becomes buoyant at $\sim 10^{17}$ G, floats up and 
breaks through the surface as a sub-burst. This self-limited process would 
continue until the differential rotational energy is used up and would 
therefore continue for several seconds with substructures given by the fluid 
instabilities on a millisecond time scale.\\
We have further analyzed the neutrino signal expected from the event. The 
neutrino emission is dominated by anti-neutrinos produced in positron 
captures, 
the total luminosities lie around $\sim 10^{53}$ erg/s and are substantially
lower than those found in previous investigations. Preliminary analysis of
the neutrino anti-neutrino pair annihilation efficiency suggests that it is
difficult to power energetic bursts with energies $\gg 10^{49}$ ergs via this 
mechanism. 
Whether this energy is enough to power a short GRB will have to be clarified
by future afterglow observations.

\begin{theacknowledgments}
The calculations reported above were performed on the United Kingdom
Astrophysical Fluids Facility (UKAFF) and on the supercomputer of the
Math Modelling Center of the University of Leicester.
\end{theacknowledgments}


\begin{thebibliography}{99}
\bibitem[1]{baumgarte00}
Baumgarte T.,  Shapiro S.,    Shibata M.,  2000, ApJ, 528, L29

\bibitem[2]{belczynski01}
Belczynski, K., Bulik, T. and Rudak, B., astro-ph/0112122(2001)

\bibitem[3]{bildsten92}
Bildsten,L. and Cutler,C., ApJ, 400, 175 (1992)

\bibitem[4]{bloom02}
Bloom, J.S, Kulkarni, S.R. and Djorgoski, S.G., AJ, 123, 1111 (2002)

\bibitem[5]{cardall01}
Cardall, C, et al. ApJ, 554, 322 (2001)

\bibitem[6]{duncan92}
Duncan, R. and Thompson, C., ApJ, 392, L9 (1992)

\bibitem[7]{eichler89}
Eichler.D. et al., Nature, 340, 126 (1989)

\bibitem[8]{frail01}
Frail, D. et al., ApJ, 562, L55 (2001)

\bibitem[9] {kluzniak98}
Kluzniak, W. and Ruderman, M., ApJ, 505, L113 (1998)

\bibitem[10]{kochanek92}
Kochanek, C.S., ApJ, 398, 234 (1992)

\bibitem[11]{narayan92}
Narayan, R., Paczynski, B. and Piran, T., ApJ, 395, L83 (1992)

\bibitem[12]{ostriker68}
Ostriker J.,  Bodenheimer P.,  1968, ApJ, 151, 1089

\bibitem[13]{prakash95}
Prakash, M. et al., Phys. Rev D52, 661 (1995)

\bibitem[14]{rosswog00}
Rosswog, S., Davies, M.B., Thielemann, F.-K. and Piran, T.,
A\&A 360, 171  (2000)

\bibitem[15]{rosswog01a}
Rosswog, S. and Davies, M.B., MNRAS in press (2002)

\bibitem[16]{rosswog01b}
Rosswog, S. et al., in preparation (2002)

\bibitem[17]{ruffert01}
Ruffert, M. and Janka, T., A\&A, 380, 544 (2001)

\bibitem[18]{shapiro00}
Shapiro, S.L., ApJ, 544, 397 (2000)

\bibitem[19]{shen98a}
Shen, H. et al., Nucl. Phys. A637, 435 (1998)

\bibitem[20]{shen98b}
Shen, H. et al., Prog. Theor. Phys. 100, 1013 (1998)

\bibitem[21]{usov92} Usov, V.V., Nature, 357, 472 (1992)

\bibitem[22]{usov94} Usov, V.V., MNRAS, 267, 1035 (1994)
\end{thebibliography}
\end{document}